\documentclass[aps,12pt]{revtex4}
\usepackage{epsfig}
\usepackage{graphics}
\usepackage{graphicx}
\usepackage{amsmath}
\usepackage{amsfonts}
\usepackage{amssymb}

\newcommand{\be}{\begin{equation}}
\newcommand{\ee}{\end{equation}}
\newcommand{\bel}[1]{\begin{equation}\label{#1}}
\newcommand{\bea}{\begin{eqnarray}}
\newcommand{\eea}{\end{eqnarray}}
\newcommand{\ba}{\begin{array}}
\newcommand{\ea}{\end{array}}

\newcommand{\bfx}{\mathbf{x}}

\def\Z{{\mathbb Z}}

\begin{document}

\setlength{\unitlength}{1mm}

\title[TASEP with sublattice parallel update]{Green functions for the TASEP \\ with sublattice parallel update}

\author{S.S. Poghosyan}
\email{spoghos@theor.jinr.ru}

\author{V.B. Priezzhev}
\email{priezzvb@theor.jinr.ru}

\affiliation{ Bogoliubov Laboratory of Theoretical Physics,
\\ Joint Institute for Nuclear Research, 141980 Dubna, Russia}

\author{G.M. Sch{\"u}tz}
\email{g.schuetz@fz-juelich.de}

\affiliation{Institut f\"ur Festk\"orperforschung, Forschungszentrum J\"ulich,
52428 J\"ulich, Germany}

\bigskip

\begin{abstract}
We consider the totally asymmetric simple exclusion process (TASEP) in discrete time with the sublattice parallel
dynamics describing particles moving to the right on the one-dimensional infinite chain with equal hoping probabilities.
Using sequentially two mappings, we show that the model is equivalent to the TASEP with the backward-ordered sequential
update in the case when particles start and finish their motion not simultaneously.
The Green functions are obtained exactly in a determinant form for different initial and final conditions.
\end{abstract}

\maketitle

\section{Introduction}

The totally asymmetric simple exclusion process (TASEP)  is a stochastic interacting particle system
which serves as a paradigmatic model for nonequilibrium behaviour \cite{Spoh91,Ligg99,Gunter}.
The dynamics of this lattice gas model is characterized by the updating law. In one dimension on the integer lattice $\Z$
the most important cases of discrete-time updates are the backward-sequential, parallel and sublattice
parallel updates \cite{Raje98}. For a finite number of particles
these dynamics can be defined through a master equation of the form
\bel{1-1}
P(\mathbf{x},t+1) = \sum_{\bfx'} p_{\bfx,\bfx'}  P(\mathbf{x}',t)
\ee
where ${\bf x} = \{x_i\}$ describes the positions of the particles and $p_{\bfx,\bfx'}$ is the transition probability to
go in one time step from a configuration ${\bf x}^{'}$ to a configuration ${\bf x}$. This transition probability is different for the
various update schemes. For the backward-sequential update, each particle may take one step to the right with probability $v$
if the target site is vacant at the beginning of the time step or becomes vacant at the end of the time step (due to motion of the particle
in front). For the parallel update, the motion to the right is allowed only if the target site is vacant at the beginning of the time step.
By iterating (\ref{1-1}) one obtains the solution of the master equation for any given
initial configuration $\bfx^0$, i.e., the conditional probability to find a particle configuration $\bfx$ at time step $t$,
given that the process started from configuration $\bfx^0$.  This stochastic many-body dynamics have a  natural
interpretation in field-theoretic terms \cite{Matt98,Gunter} where specific realizations of the process correspond
to paths in the path integral representation of field theoretic quantities.
Therefore,
in analogy to the corresponding terminology in field theory,
we refer to this time-dependent conditional transition probability as the Green function.

For the first two cases, backward-sequential and parallel update,
the Green functions of transition probabilities have been found by explicit solution of
the master equations for the systems defined on an infinite lattice \cite{Shelest,Rako05,parallel}.
The Green function has a determinantal representation similar to the one first discovered for the
continuous-time definition of the process
\cite{Schu97} where particles jump independently after an exponentially distributed random time with
fixed rate 1  \cite{Spoh91,Ligg99,Gunter}. This representation allows for
a direct derivation of the current distribution \cite{Joha00,Naga04,Rako05} and has inspired a
considerable amount of subsequent detailed analysis of dynamical properties of the TASEP and related
models, see e.g. \cite{Sasa05, Sasa07,Boro07,Boro08}
and also of the ASEP where particles are allowed to jump in both directions \cite{Tracy1,Tracy2}.

The third type of discrete-time update,  sublattice parallel, was first considered in some detail in \cite{sublattice,open}
and has subsequently been studied for various applications
both analytically and numerically \cite{Raje98,anal}.
In this paper, we derive the Green function of the TASEP with sublattice parallel update which is defined as follows.

Consider the process on $\Z$, i.e. the one-dimensional infinite chain.
Each site labeled by an integer $i$ is occupied by at most one particle which can hop only to the right in a discrete time.
At the first moment of time, we look at all $(2i,2i+1)$ pairs.
In each of them if the vertex $2i+1$ is free and the site $2i$ is occupied,
the particle of the vertex $2i$ hops to the right with probability $v$ and doesn't move with probability $1-v$.
If both sites in a pair are occupied or empty or if site $2i$ is empty and site $2i+1$ occupied,
the pair remains unchanged at that time step.
At the next step of time we apply this rule of hopping to all pairs $(2i+1,2i+2)$.
Continuing, we apply the updating rule to $(2i,2i+1)$ pairs at each odd moment of time and to $(2i+1,2i+2)$
pairs at each even moment\footnote{We remark that these dynamics can be interpreted as the action
of the transfer matrix for the six vertex model on a diagonal lattice \cite{Kand90,sublattice,open,Hone97}.}.

\section{The equivalence of the TASEP with sublattice parallel and the backward sequential updates}

\begin{figure}[tbp]
 \unitlength=1mm \makebox(60,60)[cc]
 {\psfig{file=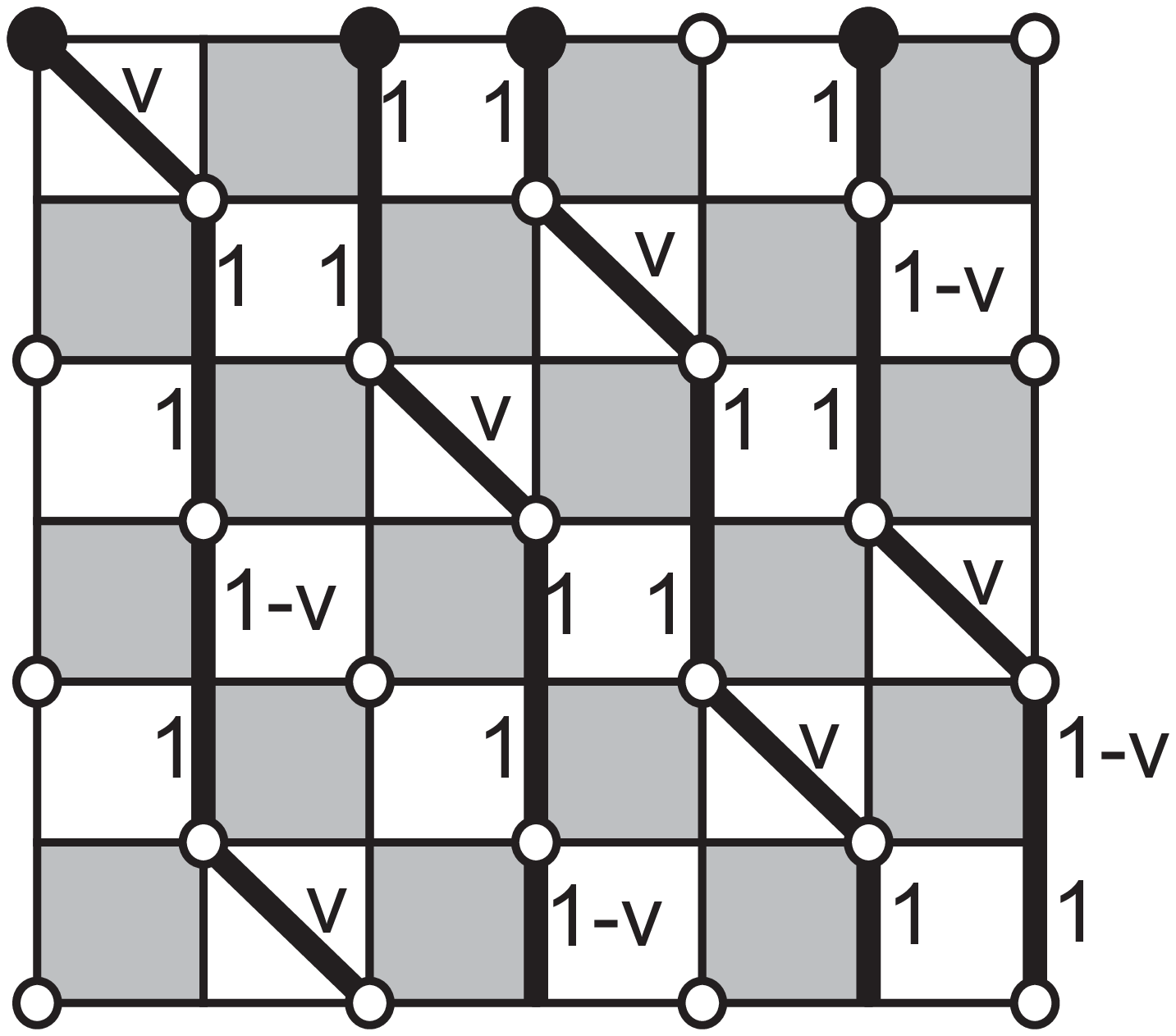,width=60mm}}
 \caption{Space-time trajectories of four particles with appropriate weights on a chessboard.}
 \label{Fig1}
 \end{figure}

As noted above, the conditional probability to find $N$ particles at positions  $x_1 < x_2 < \ldots < x_N$
at discrete time $t$ if these are in positions  $x_1^0 < x_2^0 < \ldots < x_N^0$
at initial moment of time is called the Green function of the process.
The discrete space-time dynamics can be described by a set of  trajectories on a triangle lattice
which is obtained from the square lattice by adding a diagonal bond between the upper left and the
lower right corners of each elementary square.
Being occupied by an trajectory, diagonal bonds have a statistical weight $v$ and vertical ones can have weights $1$ or $1-v$.
It is convenient to draw trajectories of particles on a chessboard  (Fig. \ref{Fig1}), where black rounds show initial positions of particles.
We notice that diagonal bonds of trajectories can be located only on white squares.
If we select a sublattice which contains upper left and lower right sites of white squares of the chessboard denoted by
white circles in Fig. \ref{Fig1}, we can see that particles effectively move on the sublattice of white vertices.
There are some exceptions at the start and at the end of trajectories.
Then, we have to consider four different cases to find a generalized determinant formula of the Green function.

\begin{figure}[tbp]
 \unitlength=1mm \makebox(150,95)[cc]
 {\psfig{file=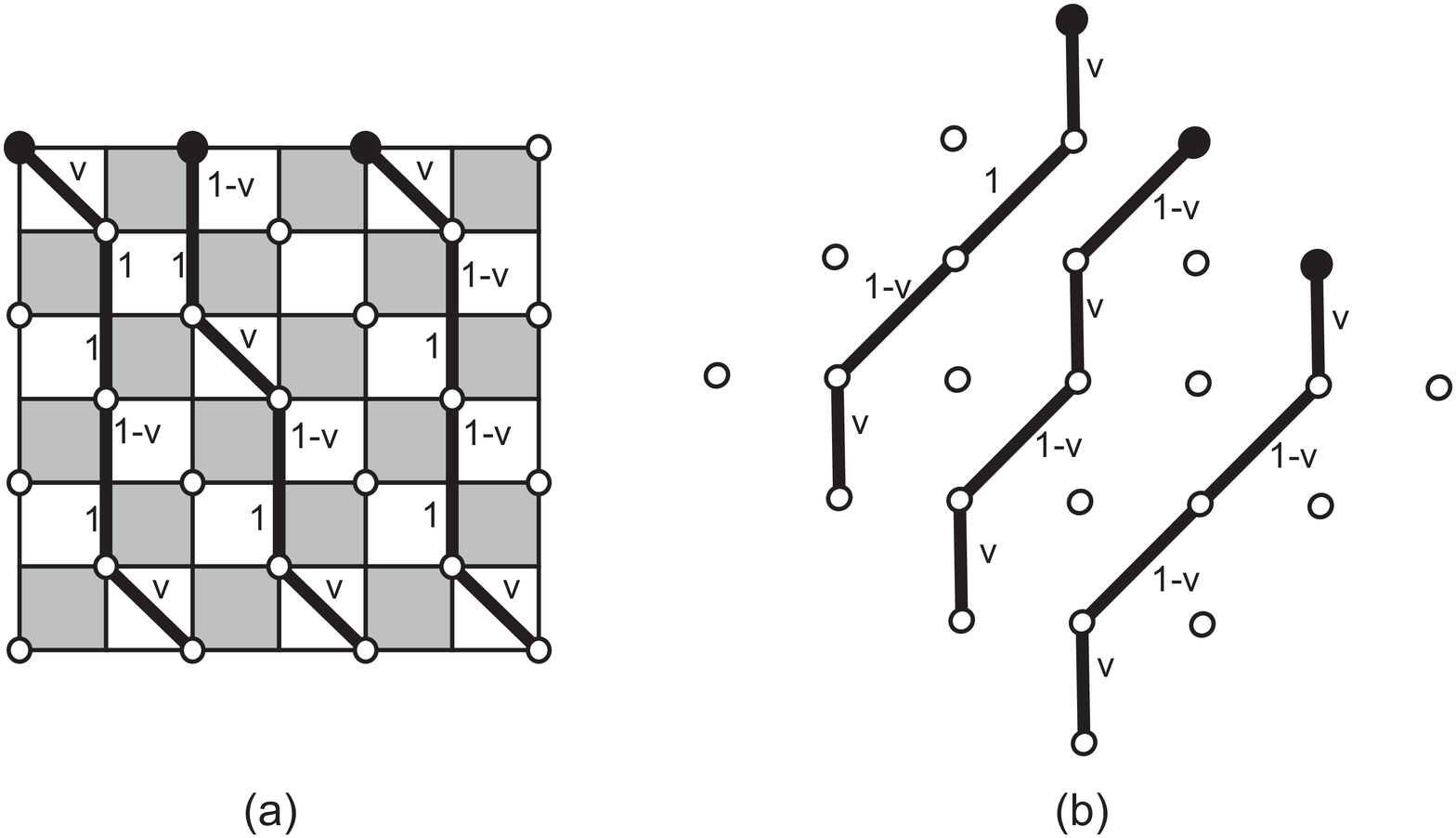,width=150mm}}
 \caption{Space-time trajectories of particles starting and ending on the sublattice of white vertices
(a) and rotated trajectories by 45 degrees clockwise (b). The time axis on the rotated lattice is directed down.}
 \label{Fig2}
 \end{figure}

Consider first the case when space-time trajectories of $N$ particles start and end on the sublattice with white vertices, Fig. \ref{Fig2}a
(the case of arbitrary initial conditions will be considered in the next section).
If we choose initial points on the white vertices of the first row,
coordinates of particles $\{x_i^0\}$ at initial times $\{T_i^0=0\}\;(i=1,2,\ldots,N)$ are even.

\begin{figure}[tbp]
 \unitlength=1mm \makebox(60,70)[cc]
 {\psfig{file=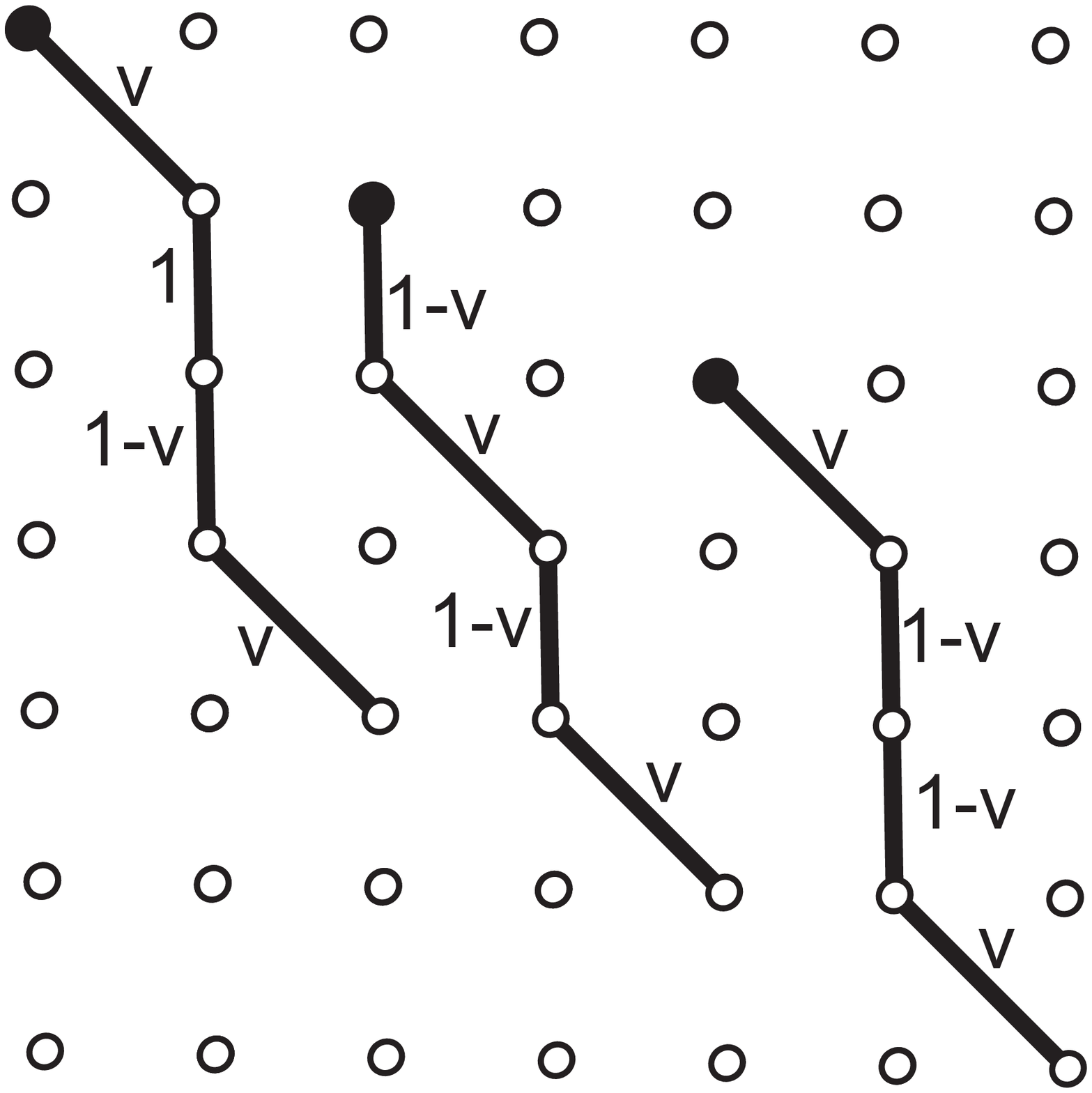,width=60mm}}
 \caption{Final worldlines of particles after appropriate transformations.}
 \label{Fig3}
 \end{figure}

The first transformation we use is a rotation of the set of trajectories by $\pi/4$ clockwise around the initial point  $\{x=0,\; T=0\}$.
Considering the vertical axis as the new time coordinate, we obtain a new set of trajectories (Fig. \ref{Fig2}b).
This set represents a new discrete-time process on the square lattice with the unit time and space intervals corresponding to vertical and
horizontal distances between neighboring sites. Starting points  change their space-time coordinates as
\begin{eqnarray}
&T_i'^0&=x_i^0/2,\\
\nonumber &x_i'^0&=x_i^0/2.
\label{rotation}
\end{eqnarray}
Now vertical bonds have weights $v$ and diagonal ones have weights $1$ or $1-v$.
We want to map them to the space-time paths of particles of the TASEP with backward-sequential update.
To this end, we shift the coordinates in each row with respect to the previous above row by $i\rightarrow i+1$.
Due to the second transformation, vertical and diagonal bonds are interchanged as it is shown in Fig. \ref{Fig3}.
The transformation of coordinates (in new units) can be written as
\begin{equation}
(x'',T'')=(x,\frac{x+T}{2}).
\label{second_trans}
\end{equation}

From Fig. \ref{Fig3} we see that  worldlines on the transformed lattice represent trajectories of particles of the TASEP
with backward-sequential update with initial space-time coordinates $\{x_i''^0,T_i''^0\}$  and final coordinates $\{x_i'',T_i''\},\; i=1,2,\ldots,N$.
The transition probability from space-time coordinates $\{x_i''^0,T_i''^0\}$ to $\{x_i'',T_i''\}$
is given by generalized determinant formula \cite{Shelest}
\begin{equation}
P\left(x_1'',T_1'';x_2'',T_2'';\ldots;x_N'',T_N''|x_1''^0,T_1''^0;x_2''^0,T_2''^0;\ldots;x_N''^0,T_N''^0\right)=\det M^{(N)},
\label{determ}
\end{equation}
where the matrix elements of $N\times N$ matrix $M^{(N)}$ are
\begin{equation}
M_{ij}^{(N)}=F_{i-j}\left(x_i''-x_j''^0,T_i''-T_j''^0\right),
\label{matelem}
\end{equation}
with the function $F_{m}\left(x,T\right)$ introduced in \cite{Shelest}
\begin{equation}
F_{m}(x,T)=\frac{1}{2\pi
 i}\int_{|z|=1-0}dz(1-v+\frac{v}{z})^T(1-z)^{-m}z^{x-1}.
\label{ffunc}
\end{equation}
Substituting transformation (\ref{second_trans}) to the determinant formula (\ref{determ})
we obtain the Green function of the TASEP with sublattice parallel update
\begin{equation}
P\left(x_1,x_2,\ldots,x_N|x_1^0,x_2^0,\ldots,x_N^0;T\right)=\det M^{(N)}
\label{determ2}
\end{equation}
with matrix elements
\begin{equation}
M_{ij}^{(N)}=F_{i-j}\left(x_i-x_j^0,\frac{T+x_i}{2}-\frac{x_j^0}{2}\right).
\end{equation}

\begin{figure}[h!]
\includegraphics[width=150mm]{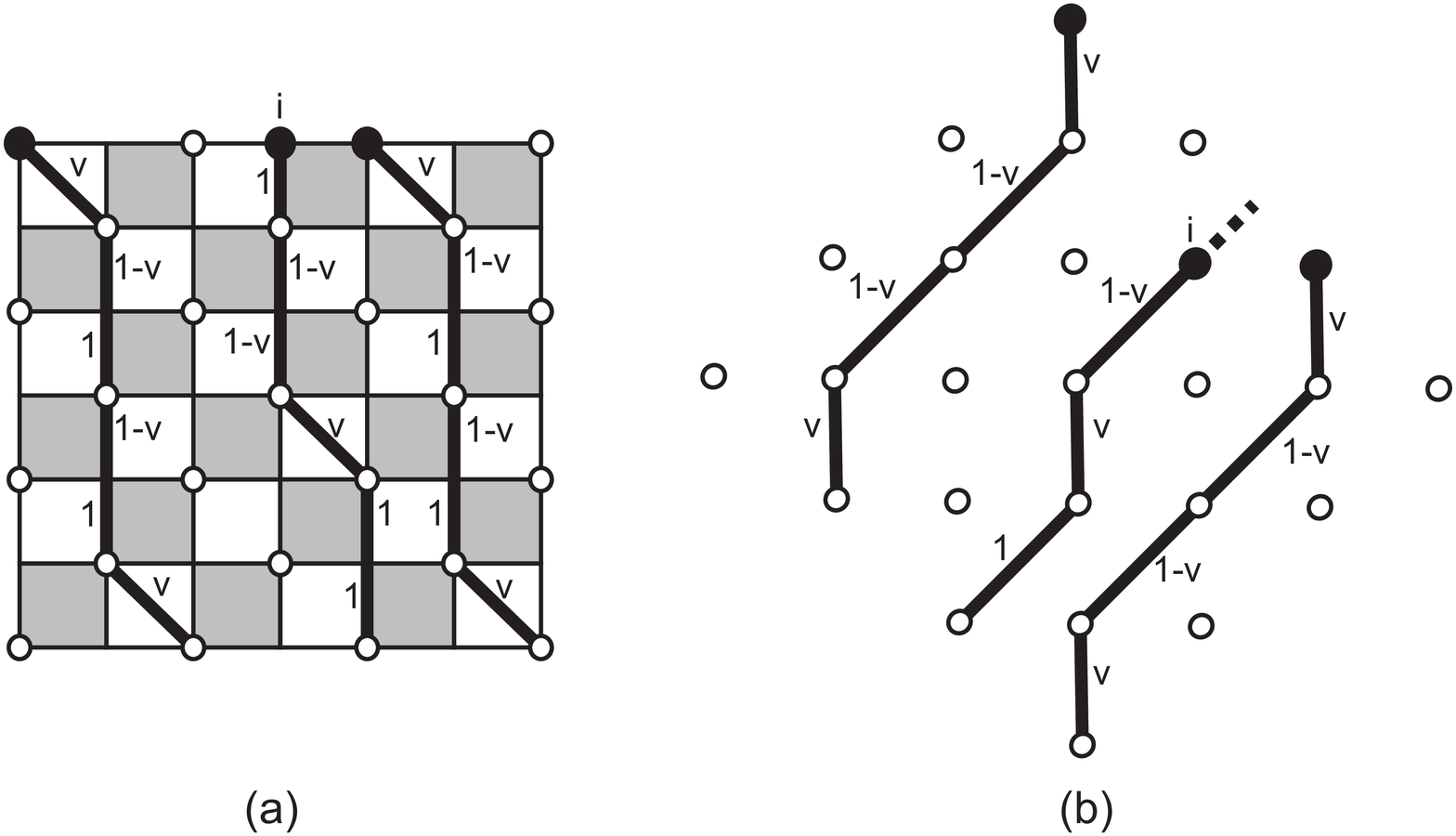}
\caption{\label{Fig4} Space-time trajectories of particles in the case,
when the $i$-th worldline starts from the non-white vertex and ends on a white one
(a) and rotated version of trajectories by 45 degrees clockwise (b).}
\end{figure}

\section{Other cases of starting and ending points }

Consider the case when the $i$-th particle starts its motion from an odd
site and ends on an even one (Fig. \ref{Fig4}a).

As the weight of the first bond of the $i$-th trajectory is $1$,
we can set the beginning of motion at the nearest white (sublattice) site and then rotate the sublattice
as in the previous case (Fig. \ref{Fig4}b).
Applying the shift transformation, we obtain for the initial coordinates of  $i$-th  particle:

\begin{equation}
(x_i^{''0},T_i^{''0})=(x_i^0,\lceil x_i^0/2\rceil)
\label{second_case}
\end{equation}
where $\lceil x\rceil$ is the ceiling function.
Substituting expressions of all starting and end points into the determinant formula ({\ref{determ}),
we derive the Green function for this case.

The third case is when the $i$-th trajectory starts from the white (even) vertex and ends on a non-white one.
We see, that if we add an additional vertical bond with weight $1$ to the last node of that trajectory,
the total weight of the whole path will not be changed. Then we can set the endpoint of $i$-th particle
at  time $T+1$. Repeating two transformations, we obtain for coordinates of the end point:
\begin{equation}
(x''_i,T''_i)=(x_i,\lceil \frac{T+x_i}{2}\rceil)
\label{third_case}
\end{equation}

The last case when the $i$-th trajectory starts and ends on non-white vertices is an obvious combination of the second and
the third case.

Generalizing all four cases of boundary conditions,
we derive following transformations for the initial and final coordinates for all types of trajectories
\begin{eqnarray}
\nonumber &T_i''^0&=\lceil x_i^0/2 \rceil,\\
\nonumber &x_i''^0&=x_i^0,\\
\nonumber &T_i''&=\lceil \frac{T+x_i}{2}\rceil,\\
&x_i''&=x_i.
\label{generaltransform}
\end{eqnarray}

Substituting new coordinates (\ref{generaltransform}) into the determinant formula (\ref{determ})
we obtain the Green function of the TASEP with sublattice parallel update
\begin{equation}
P\left(x_1,x_2,\ldots,x_N|x_1^0,x_2^0,\ldots,x_N^0;T\right)=\det M^{(N)},
\label{determ2}
\end{equation}
with matrix elements
\begin{equation}
M_{ij}^{(N)}=F_{i-j}\left(x_i-x_j^0,\lceil \frac{T+x_i}{2}\ \rceil-\lceil\frac{x_j^0}{2}\rceil\right).
\end{equation}

\section{Discussion}

Having explicit determinant expressions for the Green function, we can compare their relative advantages
and disadvantages for the three basic updates, the backward-sequential, parallel and sublattice-parallel one.
Criteria for the comparison follow from practical use of the Green function in probabilistic calculations.
To find a probability distribution for a selected particle or a correlation function for several particles in the
TASEP, detailed information contained in function $P\left(x_1,x_2,\ldots,x_N|x_1^0,x_2^0,\ldots,x_N^0;T\right)$
should be reduced by summation over a part of the final coordinates $\{x_i\}$ for fixed initial coordinates
$\{x_i^0\}$ (see e.g. \cite{Rako05}). Then, the first criterion for the comparison is simplicity of the summation
procedure  in different cases. The second criterion is simplicity of the matrix $M^{(N)}_{ij}$ itself, because
asymptotic calculations for large $N$ and $T$ need an elaborated analysis of resulting determinant expressions
(see e.g. \cite{Boro07,Boro08}). The third criterion is the presence or lack of particle-hole symmetry,
which is essential for the derivation of single-particle probability distributions in some
particular cases \cite{Rako05}.

(A) {\it The backward-sequential update.} The form of the matrix elements $M^{(N)}_{ij}$ in this case is especially
simple
\begin{equation}
M^{(N)}_{ij}=F_{i-j}(x_i-x_j^0,T)
\label{BSU}
\end{equation}
where function $F_m(x,T)$ is given by Eq.(\ref{ffunc}). The Green function $P\left(x_1,x_2,\ldots,x_N|x_1^0,x_2^0,\ldots,x_N^0;T\right)$
is uniform in variables $\{x_i\}$, so the summation procedure is straightforward \cite{Rako05}. A shortcoming of this update is lack of
the particle-hole symmetry. Indeed, due to possible transitions for one time step $x_i \rightarrow x_i+1,x_{i+1}=x_i+1 \rightarrow x_{i+1}+1,
\dots, x_{i+k}=x_{i+k-1}+1 \rightarrow x_{i+k}+1$, a hole can move in the opposite direction by jumps of length $k>1$.

(B){\it The parallel update}. The form of the matrix $M^{(N)}_{ij}$ in this case is more complicated \cite{parallel}:

\begin{equation}
M^{(N)}_{ij}=\tilde{F}_{i-j}(x_i-x_j^0,T)
\label{PAR}
\end{equation}
where
\begin{equation}
\tilde{F}_{\pm m}(N,T)=\sum_{n=0}^m\sum_{k=-n}^{\infty}(\pm 1)^n\frac{m(m+k+n-1)!}{(k+n)!n!(m-n)!}(\pm \frac{v}{1-v})^n F_0(N\pm k,T)
\label{PARffunct}
\end{equation}

The  Green function obeys the particle-hole symmetry, but a drawback is in the determinant formula
\begin{equation}
P\left(x_1,x_2,\ldots,x_N|x_1^0,x_2^0,\ldots,x_N^0;T\right)=(1-v)^n \det M^{(N)}
\label{PAR-Green}
\end{equation}
which depends on the number of pairs $n$ of neighboring particles in the final configuration. Therefore, the sum over $\{ x_i\}$ splits
into groups by the number of clusters of connected particles.

(C) {\it The sublattice parallel update.} The Green function for this update is free of shortcomings of two previous cases. It is uniform
in variables $\{x_i\}$, obeys the particle-hole symmetry and has a relatively simple analytical form (\ref{final}). A fee for this
advantage is a rather complicated time dependence in (\ref{final}) which involves both initial and final coordinates and the ceiling
function $\lceil x \rceil$. Thus, we may conclude that a proper choice of the discrete time Green function strongly depends on
peculiarities of the corresponding probabilistic problem.

\section*{Acknowledgments}
This work was supported by the RFBR grants 07-02-91561-a, 09-01-00271-a and the DFG grant 436 RUS 113/909/0-1(R).

\end{document}